\documentclass[lettersize,journal]{IEEEtran}
\newcommand{\taco}[1]{\textcolor{black}{#1}} 

\usepackage{algorithm}
\usepackage{algpseudocode}
\makeatletter
\def\BState{\State\hskip-\ALG@thistlm}
\makeatother
\usepackage{amsmath}
\usepackage{multirow}
\usepackage{multicol}
\usepackage{array}
\usepackage{subfig}
\usepackage{caption}
\usepackage{balance}
\usepackage{fancyhdr}
\usepackage{hyperref}
\usepackage[absolute]{textpos}
\usepackage{listings}
\usepackage{xcolor}
\usepackage{hyphenat}
\usepackage{multirow}
\usepackage{makecell}
\usepackage{siunitx}
\usepackage{graphicx}
\newcolumntype{?}{!{\vrule width 1pt}}
\definecolor{codegreen}{rgb}{0,0.6,0}
\definecolor{codegray}{rgb}{0.5,0.5,0.5}
\definecolor{codepurple}{rgb}{0.58,0,0.82}
\definecolor{codeblue}{rgb}{0,0,0.92}
\definecolor{backcolour}{rgb}{0.97,0.97,0.95}

\lstdefinestyle{mystyle}{
    backgroundcolor=\color{backcolour},   
    commentstyle=\color{codegreen},
    keywordstyle=\color{blue},
    numberstyle=\tiny\color{codegray},
    stringstyle=\color{red},
    identifierstyle=\color{black},
    basicstyle=\ttfamily\scriptsize,
    emph={int,char,double,float,unsigned,void,bool,assert,std,string,vector,unique_ptr,shared_ptr,LLVM_DEBUG},
    emphstyle={\color{codepurple}},
    classoffset=1, 
    otherkeywords={>,<,.,;,-,!,=,~,*,(,)},
    morekeywords={>,<,.,;,-,!,=,~,*,(,)},
    classoffset=0,
    breakatwhitespace=false,         
    breaklines=true,                 
    captionpos=b,                    
    keepspaces=true,                 
    numbers=left,                    
    numbersep=5pt,                  
    showspaces=false,                
    showstringspaces=false,
    showtabs=false,                  
    tabsize=2,
    float=t!,
    floatplacement=t!
}
\newcommand{\major}[1]{\textcolor{black}{#1}} 

\IEEEaftertitletext{\vspace{-2\baselineskip}}
\IEEEoverridecommandlockouts

\begin{document}

\title{ACPO: An AI-Enabled Compiler Framework}

\author{Amir H. Ashouri\quad Muhammad Asif Manzoor\quad Duc Minh Vu\quad Raymond Zhang\quad Colin Toft\quad \\  Ziwen Wang\quad Angel Zhang\quad Bryan Chan\quad Tomasz S. Czajkowski \quad Yaoqing Gao\vspace{1em}

Heterogeneous Compiler R\&D Lab\\ Huawei Technologies Canada \thanks{* ACPO LLVM source code: \url{https://gitee.com/openeuler/llvm-project}} 
\thanks{* ACPO ML Models: \url{https://github.com/Huawei-CPLLab/ACPO}}}





\maketitle
\begin{textblock}{20}(1.6,0.5)
\noindent * The short version is accepted at ACM/IEEE CASES'24 -- \url{https://ieeexplore.ieee.org/document/10740757} 
\\
* LLVM-Dev 2023 Presentation: \url{https://youtu.be/jzT-tar_X0U}
\\
\end{textblock}
\fancyhead{}

\begin{abstract}
The key to the performance optimization of a program is to decide correctly when a compiler should apply a certain transformation. 
This paper presents ACPO: An AI-Enabled Compiler Framework, a novel framework that provides LLVM with simple and comprehensive tools to benefit from employing ML models for different optimization passes. We first showcase the high-level view, class hierarchy, and functionalities of ACPO and subsequently, demonstrate \taco{a couple of use cases of ACPO by ML-enabling the Loop Unroll and Function Inlining passes used in LLVM's O3. and finally, describe how ACPO can be leveraged to optimize other passes. Experimental results reveal that the ACPO model for Loop Unroll can gain on average 4\%, 3\%, 5.4\%, and 0.2\% compared to LLVM's vanilla O3 optimization when deployed on Polybench, Coral-2, CoreMark, and Graph-500, respectively. 
Furthermore, by including both Function Inlining and Loop Unroll models, ACPO can provide a combined speedup of 4.5\% on Polybench and 2.4\% on Cbench when compared with LLVM's O3, respectively.}
\end{abstract}

\section{Introduction \& Related Works}
\label{sec:Intro}

State-of-the-art compilers, e.g., LLVM, are built with sophisticated optimization pipelines that result in robust, reproducible \cite{ivie2018reproducibility}, and reliable code generation \cite{Hall2009}. The LLVM compiler framework is no exception with its multi-layer design, where passes in the mid-end can perform analyses or transformations on a code region, one after another \cite{lattner2008llvm}.  LLVM passes operate on code regions with varying levels of granularity, including loops, functions, call graphs, and modules. 
Compiler engineers have continuously improved the quality of the optimization passes in LLVM, and have derived a set of fixed-length optimization pipelines, known as standard optimization levels, i.e., \texttt{O\{1,2,3,s,z\}}, which are designed to improve the code size or performance (to varying degrees) of the programs to which they are applied. At the highest optimization level available, i.e. \texttt{O3}, more than 60 unique transformation passes exist in LLVM's pipeline \cite{Ashouri2017micomp}.

Leveraging machine learning (ML) for compilers has recently become an extensively researched topic. In many instances, the authors showed experimental results that outperformed LLVM's \texttt{O3} \cite{Ashouri2016Cobayn,Ashouri2017micomp,cummins2017deep,haj2020neurovectorizer,ashouri2022mlgoperf,Patabandi2021PredictiveDL,Patabandi2023EfficientlyLL,cummins2023large,gong2024ast}. Machine learning approaches normally rely on data generated by an autotuner \cite{ansel2014opentuner,huawei2019autotuner}, which helps explore an optimization space formed by the number of parameters in the decision set. \major{Alternatively, by the growth of applications of Reinforcement Learning (RL), as opposed to generating offline training data by an autotuner, authors have explored their action space, \texttt{online}, and during the training process of the RL agent. This is done by feeding back a \texttt{reward} value to update the agent's policy \cite{trofin2021mlgo,cummins2021compilergym,ashouri2022mlgoperf,wang2022RLcompilerGym}.} To characterize the software being compiled for the ML model, program representation methods are employed, which generally fall into three categories: using static features \cite{Fursin2011,ashouri2022mlgoperf}, using dynamic features \cite{Cavazos2006a,hoste2007microarchitecture}, and graph-based characterization \cite{Park2012,ben2018ins2vec,alon2019code2vec,cummins2021programl}. \major{In recent years, there have been considerable advancements in the space, especially in the graph-based program representation techniques \cite{guo2020graphcodebert,cummins2021programl} and token-based identifiers \cite{wang2021codet5}, which are outside the scope of this work.}
Recently, there have been a few works providing an end-to-end approach to construct ML-based heuristics for optimizing the decision-making process of the transformation passes \cite{cummins2017deep,das2020GraghColoringAMD,rotem2021profilecummins,trofin2021mlgo,cummins2021compilergym,NeuralInstructionCombinerIntel2022,ashouri2022mlgoperf}. Among these, only a subset provides an ML-guided optimization approach (MLGO) \cite{cummins2021compilergym,rotem2021profilecummins,trofin2021mlgo,ashouri2022mlgoperf} baked into a single or multiple passes to streamline the optimization process inside the pass functionality readily. The recent advancements in this area beg the question of how \textit{scalable}, \textit{user-friendly}, and \textit{comprehensive} a framework should be to provide seamless integration with existing LLVM optimization passes while maintaining a clean separation of concerns between the ML and the LLVM layers.

MLGO uses ML to advise the Inliner as to whether or not a call site should be inlined to minimize the size of the generated code. 
Later, MLGOPerf \cite{ashouri2022mlgoperf} extended MLGO to optimize inlining for performance as opposed to code size and showed efficacy on several benchmarks, e.g. SPEC CPU 2006 and Cbench.  
In this work, similar to MLGOPerf, we also focus on the hard problem of \texttt{ML performance} optimization, where the key challenge is the development of model architecture and the features used to capture key performance use cases. However, we develop a scalable, extendible, end-to-end ML-guided framework to enable easy use of ML within optimization passes in compilers such as LLVM. 
Contrary to MLGO, ACPO also provides a standalone and modular feature collection class defined in different scopes of LLVM, i.e., Module, Function, CallGraph, and Loops. Users can easily add selected features to their ACPO passes by the feature names or scopes.  
To summarize, ACPO proposes the following contributions:

\begin{enumerate}
    \item A framework to provide a comprehensive set of predefined program features, libraries, and algorithmic methods by enabling compiler engineers with a user-friendly interface to instantiate \texttt{ACPOModel} classes to replace LLVM's existing hard-coded heuristics. 
    \item ML APIs and the compiler are seamlessly connected, but at the same time, they are not interdependent; changing compiler versions, ML models, or ML frameworks won't break the functionality of ACPO. As long as the inputs/outputs match, users can easily revise the ML side without rebuilding LLVM after every change.  
    \item Showcasing the benefits of ACPO, we demonstrate two different scenarios with LLVM:\\
    a) Loop Unroll Pass --- Building both the interface and the ML model. 
    b) Function Inlining Pass --- Building the interface and leveraging an existing ML model \cite{ashouri2022mlgoperf}. 
\end{enumerate}

The rest of the paper is organized as follows. Section \ref{sec:related} discusses the state of the art. Section \ref{sec:proposed} gives details on our proposed framework and how the different layers of abstraction communicate with each other. Under Sections \ref{sec:acpolumodel} and \ref{sec:acpofimodel}, we showcase how Loop Unroll and Function Inlining passes can benefit by instantiating \texttt{ACPOLUModel}. Section \ref{sec:res} showcases the experimental results including the individual and the combined results of deploying both models and Finally, in Section \ref{sec:discussion} we discuss challenges and propose directions for future work. 

\section{Related Work}
\label{sec:related}

A survey on the compiler autotuning and optimization \cite{ashouri2018survey} reveals that there have been a vast number of approaches to employ ML techniques for automatically constructing optimization heuristics. These approaches mainly focused on two major problems of compiler optimization: (1) the optimization selection \cite{bodin1998iterative,Cooper2002, Agakov2006,fursin2009collective,Park2013,Ashouri2016Cobayn} and (2) the phase ordering problem \cite{Kulkarni2012,Ashouri2017micomp,nobrePhase2018,huang2019autophase,mammadli2020-phaseordering,cummins2021compilergym,wang2022RLcompilerGym}. \major{Further classification of the two aforementioned problems and by the type of the approach provides insights for categorizing the works into three groups: (1) autotuning-based approaches \cite{Fursin2011, chen2012deconstructing, Ashouri2013VLIW, ansel2014opentuner, huawei2019autotuner,cummins2021compilergym,wang2022automating, park2022srtuner}, (2) end-to-end frameworks of optimizations \cite{Fursin2008,cummins2017deep, haj2020neurovectorizer,cummins2021compilergym,rotem2021profilecummins,das2020GraghColoringAMD,NeuralInstructionCombinerIntel2022, ashouri2022mlgoperf,wang2022automating}, and (3) ML-guided approaches \cite{rotem2021profilecummins,trofin2021mlgo,ashouri2022mlgoperf}.}

\major{Autotuning-based approaches leverage an iterative methodology to wrap around the compiler and to speed up the search in finding better solutions to optimizing a single or multiple passes, often constructing a new sequence of optimization to outperform standard optimization levels through phase-ordering or optimization selection \cite{Agakov2006,Ashouri2016Cobayn,cummins2021compilergym,wang2022RLcompilerGym}. This can be done in a coarse \cite{Kulkarni2012,Ashouri2016Cobayn,Ashouri2017micomp} or fine-grained manner \cite{huawei2019autotuner,mangpo2021,cummins2021compilergym}. End-to-end approaches often provide a framework that can be paired or built with the compiler to offer boosted search or ordering of optimizations \cite{cummins2017deep,haj2020neurovectorizer,NeuralInstructionCombinerIntel2022}.  ML-guided approaches build the decision-making capabilities of ML models into single or multiple passes that can act as a standalone ML-enabled pass. Their corresponding ML models can be either built with the compiler using an Ahead-of-Time (AOT) compilation or be used at runtime using an alternative interprocess communication method, e.g., Python named pipes. All things considered, several recent works can be included in multiple classes, therefore, the lists here are not mutually exclusive.}

\major{Cummins et al. \cite{cummins2021compilergym} propose CompilerGym, a fine-grained RL environment to speed up the search for tuning a single or multiple compiler passes defined as action space. The authors leverage OpenAI's Gym \cite{Brockman2016OpenAIG} library to provide a flexible environment to tackle the selection and the phase-ordering of optimization. The results showcase the benefits of using their RL agent to reduce the code size of Cbench benchmarks further. Contrary to this work, we propose an ML-guided approach to provide a framework for optimizing passes to be used standalone or inside LLVM's O3 pipeline. Additionally, our objective is to improve the performance instead of reducing a binary's code size. }

Rotem and Cummins \cite{rotem2021profilecummins} propose a Decision Trees (DS)-based framework to infer branch probabilities for use in providing profile-guided optimizations (PGO). The authors use hand-crafted features for the characterization of code regions and how they are able to reap the benefits of PGO without actually running the program with instrumentation to collect profile data and recompiling. \major{This work is relevant to ours since the approach builds into LLVM and once trained, it can provide PGO  with certain profiling information without actually running PGO.}

\major{Trofin et al. \cite{trofin2021mlgo} propose MLGO that uses ML to advise the Inliner as to whether or not a call site should be inlined, in order to minimize the size of the generated code. MLGO framework compiles the ML model into C++ libraries using Ahead-of-Time (AOT) compilation which becomes part of the compiler at build time. Additionally, the authors have extended the code base to provide optimization for the register allocation problem. Later, MLGOPerf \cite{ashouri2022mlgoperf}, extended MLGO to optimize inlining for performance as opposed to code size and showed efficacy on several benchmarks, e.g. SPEC CPU 2006 and Cbench}.  This work is relevant to us as introduced and upstream the necessary libraries and foundation to train and deploy an ML model to reduce code size reduction. 

As mentioned earlier, contrary to MLGO, ACPO focuses on the hard problem of \texttt{ML performance optimization}. We show that ACPO is easily extendible and scalable to any optimization passes and we also provide a modular feature collection class, defined in different scopes of LLVM, Module, Function, CallGraph, and Loops,  so users can leverage the ready-to-use set of features in their passes of interest.

\section{Proposed Methodology}
\label{sec:proposed}

\begin{figure*}[!t]
\centering
\subfloat[Training Flow]{
\label{fig:training}
\resizebox{.45\textwidth}{!}{\includegraphics{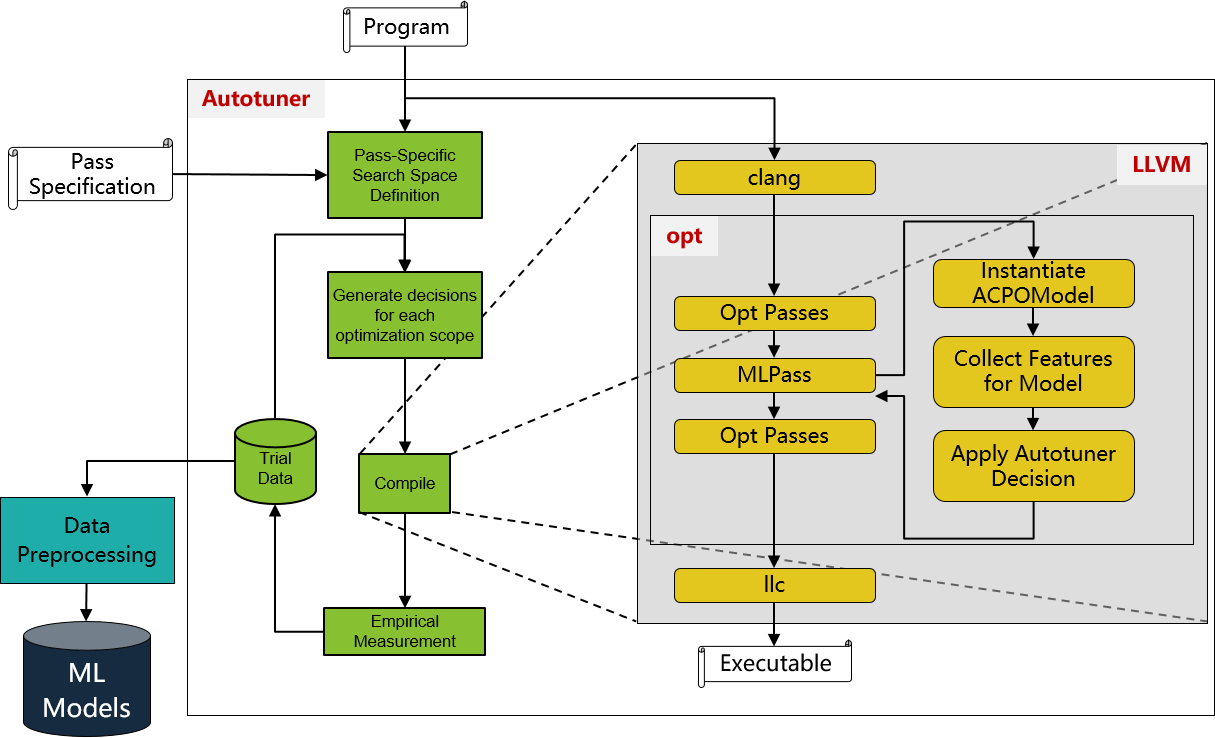}}}
\subfloat[Inference Flow]{
\label{fig:inference}
\resizebox{.47\textwidth}{!}{\includegraphics{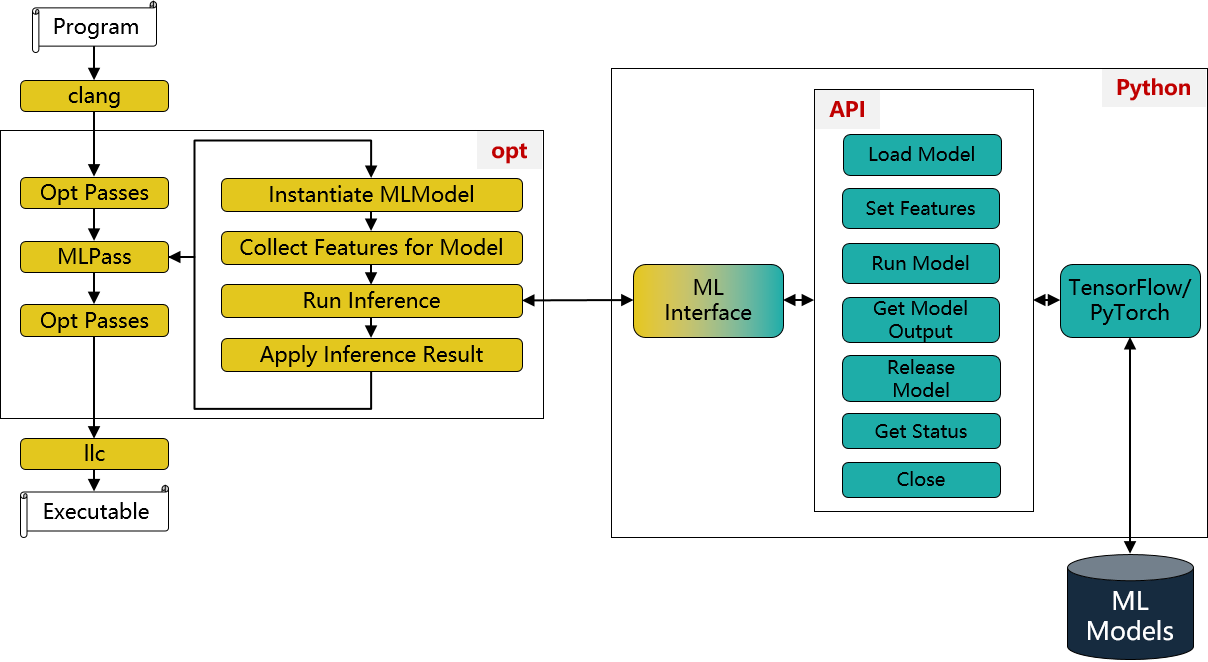}}}
\caption{ACPO Flow} 
\label{fig:acpo_training_inference}
\end{figure*}

Figure \ref{fig:acpo_training_inference} presents the high-level flow of the design of our infrastructure. There are multiple components involved with the ACPO architecture: Autotuner, compiler, \texttt{ACPOModel} abstraction, ML APIs, and ML Framework. For the most part, the autotuner is a wrapper around the compiler and its hooks in the compiler are used to generate model training data. During the inference flow, i.e., when ML models are used for decision-making within the compiler, the autotuner is not used, and we can think of the compiler as being at the top of the stack.
The ACPO inference flow begins when an ML-enabled pass is invoked and creates an instance of \texttt{ACPOModel}. The instance of \texttt{ACPOModel} aggregates appropriate features from the analyzed code region and specifies the kind of output required from the ML model. The \texttt{ACPOModel} object then transfers the features and information about the model being invoked to our ML framework via a set of remote procedure call APIs. The APIs provide the ability to send requests, including loading an appropriate model into the ML framework, executing an inference call, and returning the result of the inference back to the compiler. The compiler then applies transformations to the input program as prescribed in the inference output it receives.
In the following, we describe in more detail the training and inference flows, and then elaborate on each component of the ACPO framework to show how they fit with the training and inference flows.
\begin{figure}[!t]
\centering
\includegraphics[width=.47\textwidth]{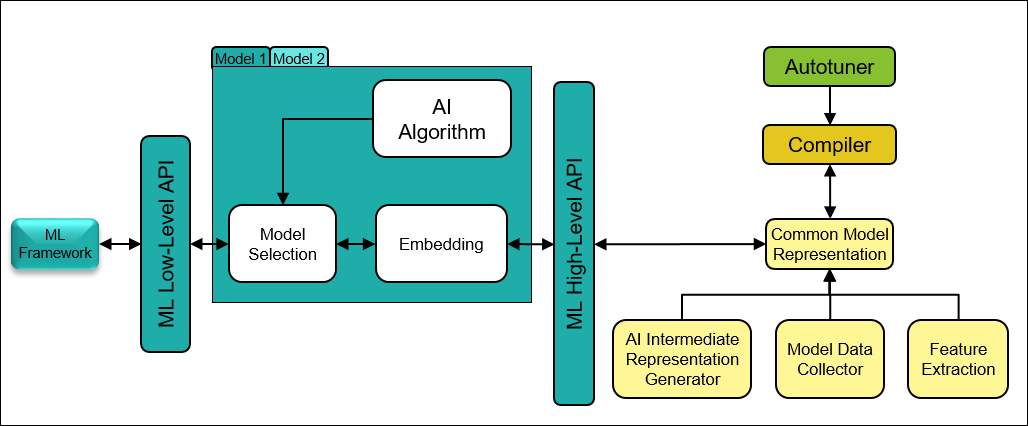}
\caption{ACPO Architecture Infrastructure} 
\label{fig:acpoInfra}
\end{figure}

\paragraph{Autotuner}
\label{sec:proposed:autotuner}

An autotuner is a software tool that derives decisions for an optimization scenario (minimization or maximization), by intelligently exploring the design space of compiler optimization parameters \cite{palermo2005multi,ansel2014opentuner,ashouri2018survey,ashouri2018book}. 
The autotuner parameters can be either coarse- or fine-grained. The former includes the selection and the phase-ordering of different optimization passes as a black box, as well as optimization parameters that apply to the whole translation unit, while the latter facilitates the tuning of optimization parameters for individual code regions of a program, e.g. instructions, loops, functions, etc., independently of each other. In this work, we perform fine-grained autotuning with an autotuner \cite{huawei2019autotuner}, which is built on top of OpenTuner \cite{ansel2014opentuner}. This allows us to generate meaningful training data and target labels that capture the effect of optimization parameters on code region types of interest. 

\subsection{ACPO Training}
\label{sec:proposed:training}

A key aspect of creating an ML-driven compiler is the training of models on sample data to ensure that the compiler makes decisions consistently with good results contained in the training data. To do this, we opted for a two-step approach: generate data based on a set of benchmarks the model is expected to reasonably reproduce and then train the model with the collected data. The data generation step is shown in Figure \ref{fig:training}.
To generate data for training purposes we rely on the autotuner. The autotuning process allows us to override compiler heuristics with an iterative search for the optimal optimization decisions for a given application. By performing the search on our benchmark programs and recording each step of the search, we can obtain information about the impact of each decision on program performance, and thus collect sufficient information to be able to train the eventual model with a reasonable chance of success. 
We use the \texttt{O3} compilation as our baseline and measure impact in terms of speedup relative to the \texttt{O3} compilation.
This process is repeated many times to gather as much data per benchmark as possible to achieve beneficial results on most benchmarks. The resulting trial data is then processed and used as input to the model training.

\subsection{ACPO Inference}
\label{sec:proposed:inference}

The compilation process involving ML inference is depicted in Figure \ref{fig:inference}. The input program is supplied to Clang to be translated into LLVM Intermediate Representation (IR). The LLVM IR is then passed through a sequence of passes and eventually reaches a pass that uses \texttt{ACPOModel} for its profitability analysis. The pass proceeds normally to confirm the legality of the transformation; when it needs to decide the profitability of the transformation, the ML framework is engaged.
The first step in the inference process is the instantiation of the ML model, via the \texttt{ACPOModel} abstraction. This step sets up the interface to the underlying ML framework, such as TensorFlow \cite{abadi2016tensorflow}. Upon successful framework initialization, \texttt{ACPOModel} collects features required for inference and sends them to the ML framework. The ML framework is then instructed to run inference on the model, and the inference result is returned to the compiler. If the inference result is such that the transformation should be applied to a given code region, the transform pass then continues and applies the transformation. Otherwise, the pass gracefully rejects the option to transform the code and proceeds to either the next code region to be analyzed or terminates. In some cases, the inference result can contain a numeric value, e.g. the predicted optimal unroll factor for a given loop.
\subsection{ACPOModel}
\label{sec:proposed:ACPOModel}

\begin{figure}[!t]
\scriptsize
\begin{lstlisting}[language=C++, label={list:ACPOModel_header}, caption=ACPOModel Header]
class ACPOModel {
public:
  ACPOModel(...);
  ~ACPOModel();
  void setMLIF(std::shared_ptr<ACPOMLInterface> ML);
  std::shared_ptr<ACPOMLInterface> getMLIF();
  void addRequiredResultField(std::string Name, Type::TypeID &ID);
  std::unique_ptr<ACPOAdvice> getAdvice();
  void addFeature(int64_t ID, Constant *Val);
  void setCustomFeatures(const std::vector<std::pair<std::string, Constant *>> &FeatureValues);

protected:
  virtual void prepareModelInput();
  virtual bool runModel(std::unique_ptr<ACPOAdvice> &Result);
};
\end{lstlisting}
\end{figure}
The \texttt{ACPOModel} class is responsible for storing model-specific information and leveraging ML APIs to perform inference. The header file code snippet is left out of this version for brevity, but it contains the same functionality described in Figure \ref{fig:acpo_training_inference}.
This class is a simple abstraction of how the model is incorporated into the compiler flow. The class is instantiated, and a reference to the ML Interface (\texttt{ACPOMLInterface} class) is provided to enable the class to communicate with the ML framework. 
The model is then configured to return a specific type of result as a part of the \texttt{ACPOAdvice} class. Then, the features to be used as input for inference are specified as input to \texttt{addFeature} method. Finally, to perform inference, the caller would invoke the \texttt{getAdvice()} method and act on the result provided.
It is worth noting that the \texttt{prepareModelInput()} and runModel() methods are virtual to enable derivation of this class to represent a new custom models.

\vspace{-1em}
\subsection{ML APIs \& Framework}
\label{sec:proposed:ml_api}

The snippet for the ACPOMLInterface header is shown under the Appendix section in Listing \ref{list:ACPOMLInterface}.

An important part of the \texttt{ACPOModel} class is the reference to the \texttt{ACPOMLInterface} class. This class is the basis for the ML API in our framework and separates the actual ML model details by abstracting the underlying ML framework away from \texttt{ACPOModel} class and the compiler. The code snippet is left out of this version for brevity, but it contains the same functionality described in figure \ref{fig:acpo_training_inference}.
\paragraph{Communication} In our implementation, we created a derived class that communicates with an ML framework using named pipes, i.e. \texttt{os.mkfifo()}. 
This approach allows us to completely separate the compiler from the ML framework, allowing the model implementation to be on the Python side and for it to be changed without modifying the compiler on a regular basis. While at the moment, the Python side of the ACPO framework is very lightweight, its separation from the compiler permits easy build-up of tools and utilities to support more models and model types without compiler changes. 
\paragraph{Feature Collection Class} We have implemented a \texttt{ACPOCollectFeature} class having around 130 handcrafted features in total at different scopes of LLVM's IR, from which 80 are Function level, 35s are Loops, and around 15 are Module features. Users can easily select features individually or by their scope when creating ACPOModel classes. 

\section{ACPO Models}
\label{sec:acpomodels}

\taco{In this section, we provide two use cases of leveraging ACPO framework on optimizing compiler passes. As we mentioned earlier, state-of-the-art approaches showcase the benefits of adapting ML models for coarse-grain and fine-grain tuning of a number of compiler optimization passes 
We speculate, ACPO can benefit any mid/back-end optimization/transformation pass with an internal decision process mechanism and a set of inputs/outputs. Historically, compiler developers have incrementally improved such optimization passes by means of algorithmic and heuristic approaches \cite{aho1986compilers,Hall2009} and we believe leveraging ML to enhance the internal decision-making process of optimization passes merits an in-depth investigation and lays the groundwork for many future works. To this end, we choose \texttt{Loop Unroll} and \texttt{Function Inlining} as two of the most important compiler optimizations used in many state-of-the-art compilers, including LLVM.}

\subsection{ACPO Loop Unrolling Model}
\label{sec:acpolumodel}

Identifying the optimal unrolling value by which a loop needs to be unrolled is known to be a hard problem with a non-polynomial complexity \cite{sarkar2000unrolling,ashouri2018survey}. There have been multiple works trying to automatically construct heuristics to find the optimal unrolling count \cite{sarkar2000unrolling,lokuciejewski2009combining,murthy2010optimal,zacharopoulos2018machine}.

The snippet for the ACPOLUModel is shown under the Appendix section in Listing \ref{list:ACPOLUModel}
A new ACPOModel class, namely \texttt{ACPOLUModel} can be instantiated as shown under Listing \ref{list:ACPOLUModel}. 
To leverage ML in unrolling, we invoked the \texttt{ACPOLUModel} after the loop unrolling legality checks had been made. Also, we respected the high-priority overrides. The highest level is set by the explicit use of unroll-count by the user, followed by the usage of \texttt{\#pragma clang loop unroll\_count(N)}.
We invoke ACPOLUModel object subsequent to these two priorities as shown in Listing \ref{list:ACPOLUModel_invocation}.

\begin{figure}[!t]
\footnotesize
\begin{lstlisting}[language=C++, label={list:ACPOLUModel_invocation}, caption=ACPOLUModel Invocation]
  if (EnableACPOLU.getNumOccurrences() > 0 && LAA) {
    std::unique_ptr<ACPOLUModel> LU = std::make_unique<ACPOLUModel>(L, ORE);
    ...
    LU->setCustomFeatures(...);
    ...
    std::unique_ptr<ACPOAdvice> Advice = LU->getAdvice();
    Constant *ValType = Advice->getField("LU-Type"); 
    Constant *Val = Advice->getField("LU-Count");
    ConstantInt *IntVal = dyn_cast<ConstantInt>(Val);
    ConstantInt *LUType = dyn_cast<ConstantInt>(ValType);
    ...
  }
\end{lstlisting}
\end{figure}

In the code above we instantiate an \texttt{ACPOLUModel} instance, which in itself creates a persistent ML interface instance to a Python-based ML Framework. The persistent interface is simply a parallel process that is launched when inference is required. However, unlike invoking blocking calls, the ML interface remains live as a separate process with which the compiler can communicate. In the case of a subsequent loop unrolling event, the process is not created anew. Instead, a reference to an existing interface is returned and the loop unrolling pass communicates with the same ML framework instance as in the previous inference step. This permits us to keep models live in memory through the compilation process. This is particularly useful in cases such as unrolling, which is often used more than once in a compiler pipeline. This way we do not have to pay the compile-time cost of restarting and closing the ML framework, or loading ML models into memory many times. 

Once the \texttt{ACPOLUModel} is instantiated, we invoke setCustomFeatures() to specify features as inputs to the model. This fulfills the requirements of the \texttt{ACPOLUModel} class and ensures all inputs are provided as needed before an inference is called. The \texttt{getAdvice} method is then called, which in turn invokes \texttt{getAdviceML} which facilitates the process of inference on behalf of the compiler. The result is then decomposed into unroll-type and unroll-factor fields, which are then obeyed by the unrolling pass and applied as directed. Note that in this work we only replace the profitability checks with ML models. Decisions regarding transformation legality are left with LLVM's legality checks in a given pass, and if those checks fail, the transformation will not proceed, nor will the profitability model be invoked.

\subsection{ACPO Function Inlining }
\label{sec:acpofimodel}

\taco{In this section, we show how to leverage ACPO to instantiate ACPOModel class for function inlining. Function inlining is one of the fundamental compiler optimizations used by many state-of-the-art compilers. In recent years, several attempts have been made to leverage ML on optimizing the pass for both code size reduction \cite{trofin2021mlgo,theodoridis2022inlining} and performance \cite{ashouri2022mlgoperf} and it was shown the optimal function inlining problem has an exponential increase in optimization space \cite{ashouri2018survey,theodoridis2022inlining}. A new ACPOModel class, namely, \texttt{ACPOFIModel} can be instantiated as shown under Listing \ref{list:ACPOFIModel} .}

The snippet for the ACPOLFIModel is shown under the Appendix section in Listing \ref{list:ACPOFIModel}

\taco{We invoked the \texttt{ACPOFIModel} by adding getACPOAdvice() after the legality checks have been made. Also, we respected the high-priority overrides. We invoke ACPOFIModel object and provide the inlining information to the pass as shown in Listing \ref{list:ACPOFIModel_invocation}}. 

\begin{figure}[!t]
\footnotesize
\begin{lstlisting}[language=C++, label={list:ACPOFIModel_invocation}, caption=ACPOFIModel Invocation]
 /// helper function for getting advice with acpo infrastructure
 bool getACPOAdvice(CallBase *CB, std::unique_ptr<ACPOFIModel> &FI,
                   ModelDataFICollector *MDC, InlineAdvisor *Advisor) {
  bool ShouldInline = false;
  if (EnableACPOFI.getNumOccurrences() > 0) {
    std::unique_ptr<ACPOFIModel> FI = std::make_unique<ACPOFIModel>(*CB, ORE);
    ...
    FI->setCustomFeatures(...);
    ...
    std::unique_ptr<ACPOAdvice> Advice = FI->getAdvice();
    Constant *ValType = Advice->getField("FI-ShouldInline"); 
    ConstantInt *ACPOInline = dyn_cast<ConstantInt>(Val);
    ShouldInline = ACPOInline->getSExtValue();
    return ShouldInline;
  }
\end{lstlisting}
\end{figure}

\section{Experimental Evaluation} 
\label{sec:res}

Note that in an ML-guided (MLGO) approach such as ACPO, optimizing a pass overrides their existing profitability checks and thus, provides a unique opportunity to run the pass in the same predefined order the sequence of passes is run at LLVM's O3. Thus, all evaluations here are based on comparing the performance of O3 with the vanilla  or the ML-enabled passes. For the same reason and except for the performance comparison with MLGO work \cite{trofin2021mlgo} under Table \ref{tab:res:combinedCbench}, we are not evaluating ACPO with the many other aforementioned works as their method is outside the scope of an MLGO approach.  To this end, we first evaluate our two passes individually, and finally, a combined evaluation is presented. 

\subsection{ACPO Loop Unrolling Model}
\label{sec:res:lu}

\begin{figure}[!t]
\hspace{-0.68cm}
\centering
\includegraphics[trim={0 1cm 0 0},clip,width=.45\textwidth]{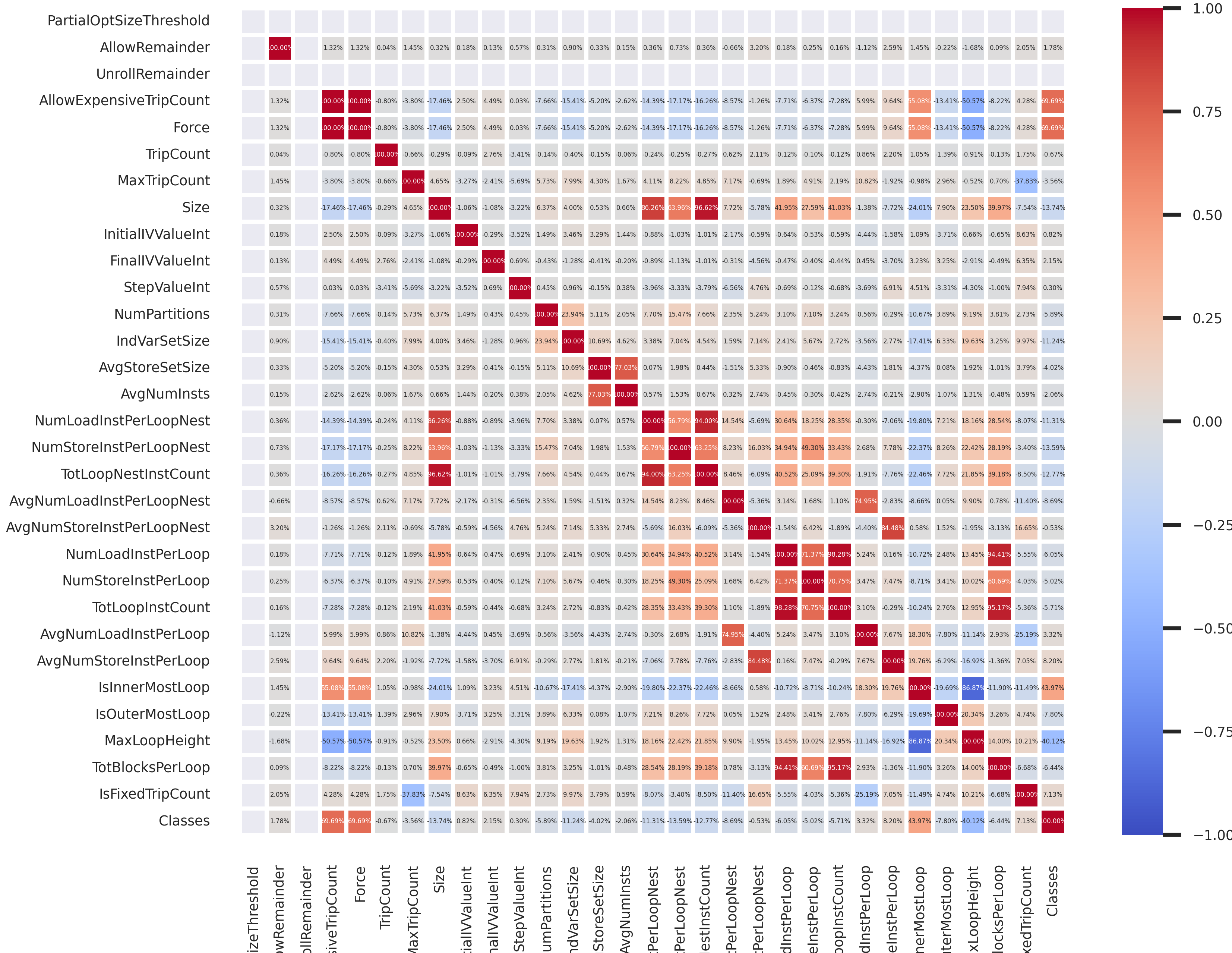}
\caption{ACPOLUModel FeaturesCorrelation Analysis} 
\label{fig:features}
\end{figure}

We design our ACPOLUModel with a 5-layer Neural Net using \texttt{ReLu} as the activation layer and a final \texttt{Softmax} layer to perform the classification task to identify the optimal unrolling factor. i.e., \texttt{UP.Count} values, given a loop fed as input. 
The input to the model is a set of 30 handcrafted features selected from the ACPOCollectFeature class at the scope of Loops. The features are passed with \texttt{std::vector} of pairs to the ML side. Additionally, the framework supports adding other types of features, i.e., graph-based, tokens, etc. However, we leverage handcrafted features derived from LLVM IR in this work. The Pearson \cite{benesty2009pearson} correlation analysis between the features and the target label, i.e. classes corresponding to unrolling count, is also depicted in Figure \ref{fig:features}.

\begin{table}[t!]
\vspace{-0.22cm}
\tiny
\centering
\caption{ACPOLUModel Features}
\begin{tabular}{|m{0.25cm}|>{\arraybackslash}m{2.85cm}|m{0.25cm}|>{\arraybackslash}m{3.15cm}|}
\hline
\textbf{No} & \textbf{Static Features} & \textbf{No} & \textbf{Static Features}  \\ \hline
1	&	PartialOptSizeThreshold      	&	16	&	NumLoadInstPerLoopNest       \\	\hline
2	&	AllowRemainder               	&	17	&	NumStoreInstPerLoopNest      \\	\hline
3	&	UnrollRemainder              	&	18	&	TotLoopNestInstCount         \\	\hline
4	&	AllowExpensiveTripCount      	&	19	&	AvgNumLoadInstPerLoopNest    \\	\hline
5	&	Force                        	&	20	&	AvgNumStoreInstPerLoopNest   \\	\hline
6	&	TripCount                    	&	21	&	NumLoadInstPerLoop           \\	\hline
7	&	MaxTripCount                 	&	22	&	NumStoreInstPerLoop          \\	\hline
8	&	Size                         	&	23	&	TotLoopInstCount             \\	\hline
9	&	InitialIVValueInt            	&	24	&	AvgNumLoadInstPerLoop        \\	\hline
10	&	FinalIVValueInt              	&	25	&	AvgNumStoreInstPerLoop       \\	\hline
11	&	StepValueInt                 	&	26	&	IsInnerMostLoop              \\	\hline
12	&	NumPartitions                	&	27	&	IsOuterMostLoop              \\	\hline
13	&	IndVarSetSize                	&	28	&	MaxLoopHeight                \\	\hline
14	&	AvgStoreSetSize              	&	29	&	TotBlocksPerLoop             \\	\hline
15	&	AvgNumInsts                  	&	30	&	IsFixedTripCount             \\	\hline
\end{tabular}
\label{tab:features}
\end{table}

\paragraph{Data Collection \& Setup} 
\label{sec:res:data}

We used an autotuner to collect a total of 100000 different unrolling configurations from Polybench by tuning the unroll count of individual loops (\texttt{UP.Count}), where the set of \texttt{UP.Count} values are $\in \{0, 2, \cdots, 64\}$. 
In this work, we define the Loop Unrolling speedup as the fraction between the base runtime and the loop runtime.
The total program runtime and the $T_{Loop}$ are collected with Linux \texttt{perf}. All the runtime measurements are done single-threaded, and we use \texttt{numactl} tool to bind the workloads to a unique CPU core. All experiments are measured on an ARMv8.2-A architecture @ 2.6GHz on Linux. We run each benchmark three times, and before each measurement run, we flush the system page cache to avoid any perturbance in the collection of our training data, and we repeat the process for all measurements having more than 1\% of variance between the runs to make sure they are robust.

\paragraph{Analysis of Compilation Time Overhead}
\label{sec:res:overehead}

We measure the compilation-time overhead of utilizing ACPOLUModel with a breakdown of each individual task, as shown in Listing \ref{list:ACPOLUModel_invocation}, and we present the breakdown in Table \ref{tab:compilation_time_overhead}. As mentioned earlier under Section \ref{sec:acpolumodel}, Loop Unroll pass is registered twice at O3's pipeline. \major{We have implemented ACPO in a way that we load our ML models once per each module and keep the models alive in the lifetime of a module to avoid a repeated loading process of the models per each instance of inference. To this end, the other contributing factor is the number of modules per application and to a lesser extent the number of loops inside a module}. The number of modules determines how many times the ML model needs to be loaded, which is an expensive task, and the number of loops determines the number of times Loop Unroll pass will be activated and ACPOLUModel will be deployed. ACPOLUModel is persistent and loaded once per module and we are working on making it even more optimized, i.e., loaded once per each application having multiple modules. For example. On Polybench's \texttt{bicg}, we have two modules, namely, \texttt{polybench.c} and \texttt{bicg.c}. The former has a single loop and the latter, has 8, thus, the overall overhead would be 0.857 seconds.

\begin{table}
\centering
\footnotesize
\caption{Compile-time Overhead Breakdown}
\label{tab:compilation_time_overhead}
\begin{tabular}{|l|c|c|}
\hline
\textbf{Task}                 & \textbf{Each Module} & \textbf{Each Loop} \\   \hline
Feature Collection            & 0.000135             & 0.000135            \\   \hline
Setting Features for ML Model & 0.000029             & 0.000029            \\   \hline
Total ML Inference Time       & 0.396148             & 0.00699             \\  \hline
Unrolling Factor Assignment   & 0.000001             & 0.000001            \\  \hline
\textbf{Total Time (s)}           & \textbf{0.396313}    & \textbf{0.007155}  \\   \hline
\end{tabular}
\end{table}

\paragraph{ML Training and Cross Validation} 
\label{sec:res:ml_cv}

The training is carried out using leave-one-out cross-validation, for which we exclude one application from the training data and use it later as the test, and this ensures no data leakage occurs.  Measured Top-1 accuracy of the \texttt{ACPOLUModel}'s predictions exhibits a geometric mean of 80\% accuracy with a standard deviation of 7\% on Polybench. 
In certain applications, e.g., trmm, 2mm, etc., the top-1 accuracy is higher than the average, and in a few such as atax, it is lower. Our investigation reveals the reason lies in the imbalanced training set for the specific cases and it can be alleviated to some extent by employing a number of preprocessing and data augmentation techniques \cite{wei2019eda}. We leave further investigation to our future works as they are outside the scope of this paper. 
\begin{table}[!t]
\centering
\caption{ACPOLUModel Top-1 Accuracy on Polybench}
\label{tab:speedup:top_1_poly}
\tiny
\begin{tabular}{|l|c|l|c|}
\hline
\textbf{Benchmark} & \textbf{Top-1 Accuracy} & \textbf{Benchmark} & \textbf{Top-1 Accuracy} \\ \hline
2mm            & 88.411 \% & gesummv        & 80 \%    \\\hline
3mm            & 84.254 \%& gramschmidt    & 75.466 \%\\\hline
adi            & 76.543 \%& lu             & 85.971 \%\\\hline
atax           & 57.5   \%& ludcmp         & 85.938 \%\\\hline
bicg           & 63.295 \%& mvt            & 75.326 \%\\\hline
cholesky       & 86.414 \%& nussinov       & 79.734 \%\\\hline
correlation    & 84.091 \%& seidel-2d      & 80.909 \%\\\hline
covariance     & 86.42  \%& symm           & 79.797 \%\\\hline
deriche        & 82.481 \%& syr2k          & 84.858 \%\\\hline
doitgen        & 75.487 \%& syrk           & 82.84  \%\\\hline
floyd-warshall & 82.864 \%& trisolv        & 69.225 \%\\\hline
gemver         & 86.547 \%& trmm           & 90.333 \%\\\hline
\multicolumn{4}{|c|}{\textbf{Geomean:\quad 80.196\%}}\\ \hline
\end{tabular}
\end{table}
Table \ref{tab:res:combinedPoly} shows the ACPOLUModel's results when deployed in a cross-validation mode. We observe that, on average, \texttt{ACPOLUModel} achieved a speedup of 4.07\% among the Polybench benchmarks with a best of 49.3\% on \texttt{Floyd-Warshall}. Further investigation reveals that the Floyd-Warshall application has a seven-loop module, which makes it for a set of 23 tunable loop-nest regions. ACPO's prediction provides for a full unroll of a couple of loops, including a multiple runtime unrolling of hot loops in the kernel. LLVM's O3, on the other hand, only unrolled two loops by their available trip count of 3 and didn't unroll the remaining hot loops. Thus, we noticed that the majority of the performance comes from deciding the correct unrolling factor for the hot loops. 
\paragraph{Generalization} To further evaluate the generalizability of the model, we have tested ACPOLUModel on a set of heavily loop-oriented benchmarks. Table \ref{tab:res:generalization} demonstrates our result for which we gain a 5.4\%  speedup over CoreMark, an average 3\% including an 18\% speedup over AMG (from Coral-2), and a 1.3\% speedup on Graph500's SSSP. Analyzing the results, we noticed that the majority of the performance comes from deciding the correct unrolling factor for the hot loops. For instance, Coral-2's AMG runtime unrolling of a hot loop located at \texttt{hypre\_CSRMatrix} Module led to an increase in performance by almost 15\%, which O3 failed to identify as profitable to unroll.

\begin{table}[!t]
\centering
\scriptsize
\caption{ACPOLUModel Generalization}
\label{tab:res:generalization}
\begin{tabular}{|l?cc?cc?c|}
\hline
\multicolumn{1}{|c?}{\multirow{2}{*}{\textbf{Benchmarks}}}               & \multicolumn{2}{c?}{\textbf{O3}}                                                                                            & \multicolumn{2}{c?}{\textbf{ACPO}}                                            & \multirow{2}{*}{\textbf{SpUp}} \\ \cline{2-5}
\multicolumn{1}{|c|}{}                                                  & \multicolumn{1}{?c|}{Time}          & Size   & \multicolumn{1}{c|}{Time}        & Size    &                                   \\ \hline
\multicolumn{1}{|l?}{CoreMark}                                          & \multicolumn{1}{l|}{20.783}        & 74328  & \multicolumn{1}{l|}{19.723}      & 74328   & \textbf{1.054}                             \\ \hline
 CLOMP              & \multicolumn{1}{l|}{142.267}       & 78952  & \multicolumn{1}{l|}{140.835}           & 78952   & \textbf{1.010}                    \\ \cline{1-6} 
Quicksilver        & \multicolumn{1}{l|}{11.647}         & 256072 & \multicolumn{1}{l|}{11.657}            & 527328  & 0.999                             \\ \cline{1-6} 
 STRIDE cac.  & \multicolumn{1}{l|}{19.707}              & 72112  & \multicolumn{1}{l|}{18.83}             & 72112   & \textbf{1.047}                     \\ \cline{1-6} 
 STRIDE str. & \multicolumn{1}{l|}{44.842}               & 71928  & \multicolumn{1}{l|}{44.792}            & 71928   & \textbf{1.001}                     \\ \cline{1-6} 
 STREAM             & \multicolumn{1}{l|}{18.899}        & 72992  & \multicolumn{1}{l|}{18.786}            & 72992   & \textbf{1.006}                  \\ \cline{1-6}  
 AMG                & \multicolumn{1}{l|}{3.2922}        & 564192 & \multicolumn{1}{l|}{2.785}             & 1292568 & \textbf{1.182}                      \\ \cline{1-6}
LAMMPS                & \multicolumn{1}{l|}{2.92}        & 3802424 & \multicolumn{1}{l|}{2.826}            & 7082568 & \textbf{1.033}                        \\ \hline
 BFS       & \multicolumn{1}{l|}{7.275}                  & 539264 & \multicolumn{1}{l|}{7.33}              & 539240  & 0.992                              \\ \cline{1-6} 
  SSSP      & \multicolumn{1}{l|}{15.508}                & 539832 & \multicolumn{1}{l|}{15.308}            & 605344  & \textbf{1.013}                             \\ \hline
\end{tabular}
\end{table}

\subsection{ACPO Function Inlining Model}
\label{sec:res:fi}



Instead of designing a novel ML model from scratch, we decided to leverage an existing model to showcase the ease of use and integration capabilities of ACPO when paired with an existing ML model. For this task, we use the MLGOPerf \cite{ashouri2022mlgoperf} model. MLGOPerf is a reinforcement learning model, and the model receives, as input, 13 function features including \texttt{block frequency}, \texttt{callsite height}, \texttt{block counts}, etc., and predicts whether or not a callsite should be inlined. 
Table \ref{tab:res:combinedCbench} showcases the experimental results of using ACPO framework for the deployment of ACPOFIModel, aka MLGOPerf's, when tested with CBench. As expected, the results exhibit similar behavior to MLGOPerf's experimental results since both leverage the same model. 

\subsection{Combined Results}
\label{sec:res:combined}


\addtolength{\tabcolsep}{-0.18em}
\begin{table}[!t]
\centering
\scriptsize
\caption{ACPO Experimental Result (Cbench)\\ Sz\%: Size bloat, Sp: Speedup. Multicolumns 1, 3, and 4 are wrt. O3 and 2, is ACPO-FI compared against MLGO \cite{trofin2021mlgo}.}
\begin{tabular}{|l|cc|cc|cc|cc|}
\hline
\multicolumn{1}{|c|}{\multirow{2}{*}{Benchmark}} & \multicolumn{2}{c|}{\textbf{ACPO-FI}}       & \multicolumn{2}{c|}{\textbf{wrt. MLGO}}     & \multicolumn{2}{c|}{\textbf{ACPO-LU}}      & \multicolumn{2}{c|}{\textbf{Combined}} \\ \cline{2-9} 
\multicolumn{1}{|c|}{}                           & \multicolumn{1}{c|}{Sz\%}        & Sp        & \multicolumn{1}{c|}{Sz\%}        & Sp        & \multicolumn{1}{c|}{Sz\%}        & Sp      & \multicolumn{1}{c|}{Sz\%}        & Sp       \\ \hline
auto\_bitcount                             & \multicolumn{1}{c|}{1}              & 0.999          & \multicolumn{1}{c|}{1}              & 1.005          & \multicolumn{1}{c|}{1.001}          & 1.003         & \multicolumn{1}{c|}{1}              & 1.001          \\ \hline
auto\_qsort1                               & \multicolumn{1}{c|}{1}              & 0.994          & \multicolumn{1}{c|}{1}              & 1.003          & \multicolumn{1}{c|}{1}              & 0.982         & \multicolumn{1}{c|}{1.91}           & 0.989          \\ \hline
auto\_susan\_c                             & \multicolumn{1}{c|}{1}              & 0.999          & \multicolumn{1}{c|}{1.567}          & 1.122          & \multicolumn{1}{c|}{3.685}          & 1.035         & \multicolumn{1}{c|}{19.797}         & 1.066          \\ \hline
auto\_susan\_e                             & \multicolumn{1}{c|}{1}              & 1.023          & \multicolumn{1}{c|}{1.567}          & 1.04           & \multicolumn{1}{c|}{3.685}          & 1.019         & \multicolumn{1}{c|}{19.797}         & 1.026          \\ \hline
auto\_susan\_s                             & \multicolumn{1}{c|}{1}              & 0.98           & \multicolumn{1}{c|}{1.567}          & 1.008          & \multicolumn{1}{c|}{3.685}          & 1.038         & \multicolumn{1}{c|}{19.797}         & 0.995          \\ \hline
bzip2d                                           & \multicolumn{1}{c|}{1.441}          & 0.991          & \multicolumn{1}{c|}{1.239}          & 1.021          & \multicolumn{1}{c|}{2.224}          & 1.01          & \multicolumn{1}{c|}{3.541}          & 1.014          \\ \hline
bzip2e                                           & \multicolumn{1}{c|}{1.441}          & 1.007          & \multicolumn{1}{c|}{1.189}          & 1.041          & \multicolumn{1}{c|}{2.35}           & 1.001         & \multicolumn{1}{c|}{3.546}          & 1.015          \\ \hline
cons\_jpeg\_c                                & \multicolumn{1}{c|}{1.288}          & 1.017          & \multicolumn{1}{c|}{1.2}            & 1.005          & \multicolumn{1}{c|}{5.385}          & 0.986         & \multicolumn{1}{c|}{5.086}          & 1.027          \\ \hline
cons\_jpeg\_d                                & \multicolumn{1}{c|}{1.289}          & 0.923          & \multicolumn{1}{c|}{1.19}           & 1.335          & \multicolumn{1}{c|}{5.4}            & 1.073         & \multicolumn{1}{c|}{4.807}          & 1.026          \\ \hline
cons\_lame                                   & \multicolumn{1}{c|}{1.437}          & 1.027          & \multicolumn{1}{c|}{1.137}          & 1.007          & \multicolumn{1}{c|}{4.501}          & 1.03          & \multicolumn{1}{c|}{4.729}          & 1.034          \\ \hline
cons\_mad                                    & \multicolumn{1}{c|}{1.217}          & 1.071          & \multicolumn{1}{c|}{1.16}           & 1              & \multicolumn{1}{c|}{4.321}          & 0.99          & \multicolumn{1}{c|}{3.193}          & 0.968          \\ \hline
cons\_tiff2bw                                & \multicolumn{1}{c|}{1.18}           & 1.002          & \multicolumn{1}{c|}{1.351}          & 1.23           & \multicolumn{1}{c|}{4.502}          & 0.943         & \multicolumn{1}{c|}{2.471}          & 0.987          \\ \hline
cons\_tiff2rgba                              & \multicolumn{1}{c|}{1.18}           & 1.25           & \multicolumn{1}{c|}{1.356}          & 1.11           & \multicolumn{1}{c|}{4.681}          & 1.036         & \multicolumn{1}{c|}{2.657}          & 1.012          \\ \hline
cons\_tiffdither                             & \multicolumn{1}{c|}{1.18}           & 1.046          & \multicolumn{1}{c|}{1.14}           & 1.014          & \multicolumn{1}{c|}{1.213}          & 0.967         & \multicolumn{1}{c|}{2.655}          & 1.002          \\ \hline
cons\_tiffmedian                             & \multicolumn{1}{c|}{1.18}           & 1.017          & \multicolumn{1}{c|}{1.048}          & 1.056          & \multicolumn{1}{c|}{1}              & 0.97          & \multicolumn{1}{c|}{3.021}          & 0.983          \\ \hline
net\_dijkstra                                & \multicolumn{1}{c|}{1}              & 0.993          & \multicolumn{1}{c|}{1}              & 0.996          & \multicolumn{1}{c|}{1}              & 0.99          & \multicolumn{1}{c|}{1}              & 1.037          \\ \hline
net\_patricia                                & \multicolumn{1}{c|}{1}              & 0.98           & \multicolumn{1}{c|}{1.052}          & 0.99           & \multicolumn{1}{c|}{1}              & 0.974         & \multicolumn{1}{c|}{1}              & 1.004          \\ \hline
off\_ghostscript                              & \multicolumn{1}{c|}{1.275}          & 1              & \multicolumn{1}{c|}{1.16}           & 1.04           & \multicolumn{1}{c|}{1}              & 1             & \multicolumn{1}{c|}{1.12}           & 1.001          \\ \hline
office\_ispell                                   & \multicolumn{1}{c|}{0.996}          & 1              & \multicolumn{1}{c|}{1.17}           & 1.02           & \multicolumn{1}{c|}{1}              & 1             & \multicolumn{1}{c|}{1}              & 1              \\ \hline
office\_rsynth                                   & \multicolumn{1}{c|}{1}              & 1              & \multicolumn{1}{c|}{1.12}           & 1.04           & \multicolumn{1}{c|}{2.024}          & 1.01          & \multicolumn{1}{c|}{2.024}          & 1.011          \\ \hline
office\_strings                           & \multicolumn{1}{c|}{1}              & 1.006          & \multicolumn{1}{c|}{1}              & 1              & \multicolumn{1}{c|}{1}              & 1             & \multicolumn{1}{c|}{1}              & 1              \\ \hline
sec\_blowfish\_d                            & \multicolumn{1}{c|}{1}              & 1.002          & \multicolumn{1}{c|}{1}              & 1.002          & \multicolumn{1}{c|}{1}              & 0.978         & \multicolumn{1}{c|}{1}              & 0.965          \\ \hline
sec\_blowfish\_e                            & \multicolumn{1}{c|}{1}              & 1.003          & \multicolumn{1}{c|}{1.39}           & 1.029          & \multicolumn{1}{c|}{1}              & 1.015         & \multicolumn{1}{c|}{1}              & 0.981          \\ \hline
security\_pgp\_d                                 & \multicolumn{1}{c|}{1.861}          & 0.993          & \multicolumn{1}{c|}{1.12}           & 1              & \multicolumn{1}{c|}{1}              & 0.991         & \multicolumn{1}{c|}{13.144}         & 1              \\ \hline
security\_pgp\_e                                 & \multicolumn{1}{c|}{1.861}          & 1.017          & \multicolumn{1}{c|}{1.12}           & 1.042          & \multicolumn{1}{c|}{1}              & 1.006         & \multicolumn{1}{c|}{13.144}         & 1.006          \\ \hline
sec\_rijndael\_d                            & \multicolumn{1}{c|}{1}              & 0.99           & \multicolumn{1}{c|}{1.39}           & 1.039          & \multicolumn{1}{c|}{3.823}          & 0.966         & \multicolumn{1}{c|}{1}              & 0.966          \\ \hline
sec\_rijndael\_e                            & \multicolumn{1}{c|}{1}              & 1.001          & \multicolumn{1}{c|}{1.09}           & 1.035          & \multicolumn{1}{c|}{3.823}          & 0.974         & \multicolumn{1}{c|}{1}              & 0.995          \\ \hline
security\_sha                                    & \multicolumn{1}{c|}{0.999}          & 1.27           & \multicolumn{1}{c|}{1.009}          & 1.132          & \multicolumn{1}{c|}{1}              & 1.174         & \multicolumn{1}{c|}{0.999}          & 1.501          \\ \hline
tele\_adpcm\_c                                & \multicolumn{1}{c|}{1}              & 1              & \multicolumn{1}{c|}{0.996}          & 1.001          & \multicolumn{1}{c|}{1}              & 1.008         & \multicolumn{1}{c|}{1}              & 1.038          \\ \hline
tele\_adpcm\_d                                & \multicolumn{1}{c|}{1}              & 0.997          & \multicolumn{1}{c|}{1}              & 1              & \multicolumn{1}{c|}{1}              & 1.132         & \multicolumn{1}{c|}{1}              & 1.138          \\ \hline
tele\_CRC32                                   & \multicolumn{1}{c|}{1}              & 1.094          & \multicolumn{1}{c|}{1}              & 1.066          & \multicolumn{1}{c|}{1}              & 1.042         & \multicolumn{1}{c|}{1}              & 1.112          \\ \hline
telecom\_gsm                                     & \multicolumn{1}{c|}{1.81}           & 1.076          & \multicolumn{1}{c|}{1.05}           & 1.04           & \multicolumn{1}{c|}{1.815}          & 1.15          & \multicolumn{1}{c|}{1.81}           & 1.117          \\ \hline
Geomean                                          & \multicolumn{1}{c|}{\textbf{1.152}} & \textbf{1.021} & \multicolumn{1}{c|}{\textbf{1.161}} & \textbf{1.041} & \multicolumn{1}{c|}{\textbf{1.818}} & \textbf{1.01} & \multicolumn{1}{c|}{\textbf{2.442}} & \textbf{1.024} \\ \hline
\end{tabular}
 \label{tab:res:combinedCbench}
\end{table}

\begin{table}[!t]
\centering
\scriptsize
\caption{ACPO Experimental Results (Polybench)}
\begin{tabular}{|l|cc|cc|}
\hline
\multicolumn{1}{|c|}{\multirow{2}{*}{Benchmark}} & \multicolumn{2}{c|}{\textbf{ACPO-LU}}                     & \multicolumn{2}{c|}{\textbf{ACPO   Combined}}        \\ \cline{2-5} 
\multicolumn{1}{|c|}{}                           & \multicolumn{1}{c|}{Size Bloat \%}       & Speedup           & \multicolumn{1}{c|}{Size Bloat \%}     & Speedup        \\ \hline
2mm                                              & \multicolumn{1}{c|}{1.903}            & 0.984             & \multicolumn{1}{c|}{1.903}          & 0.984          \\ \hline
3mm                                              & \multicolumn{1}{c|}{1.904}            & 0.983             & \multicolumn{1}{c|}{2.805}          & 0.976          \\ \hline
adi                                              & \multicolumn{1}{c|}{1.0002}           & 1.012             & \multicolumn{1}{c|}{1.38}           & 1.04           \\ \hline
atax                                             & \multicolumn{1}{c|}{0.999}            & 0.985             & \multicolumn{1}{c|}{1}              & 0.987          \\ \hline
bicg                                             & \multicolumn{1}{c|}{1}                & 1.02              & \multicolumn{1}{c|}{1}              & 1.022          \\ \hline
cholesky                                         & \multicolumn{1}{c|}{1.001}            & 1.102             & \multicolumn{1}{c|}{1}              & 1.063          \\ \hline
correlation                                      & \multicolumn{1}{c|}{1.005}            & 0.957             & \multicolumn{1}{c|}{1.901}          & 1.002          \\ \hline
covariance                                       & \multicolumn{1}{c|}{1.001}            & 1.003             & \multicolumn{1}{c|}{1.902}          & 1.009          \\ \hline
deriche                                          & \multicolumn{1}{c|}{1.0018}           & 1.078             & \multicolumn{1}{c|}{1.902}          & 1.115          \\ \hline
doitgen                                          & \multicolumn{1}{c|}{1.0001}           & 1.011             & \multicolumn{1}{c|}{1.901}          & 1.013          \\ \hline
floyd-Warshall                                   & \multicolumn{1}{c|}{1.0004}           & 1.493             & \multicolumn{1}{c|}{1}              & 0.974          \\ \hline
gemver                                           & \multicolumn{1}{c|}{1.0004}           & 0.958             & \multicolumn{1}{c|}{1.902}          & 1.494          \\ \hline
gesummv                                          & \multicolumn{1}{c|}{1}                & 1.143             & \multicolumn{1}{c|}{1}              & 0.975          \\ \hline
gramschmidt                                      & \multicolumn{1}{c|}{1}                & 0.91              & \multicolumn{1}{c|}{1}              & 0.995          \\ \hline
lu                                               & \multicolumn{1}{c|}{1.001}            & 0.978             & \multicolumn{1}{c|}{1.9}            & 1              \\ \hline
ludcmp                                           & \multicolumn{1}{c|}{1.0008}           & 1.047             & \multicolumn{1}{c|}{1}              & 0.996          \\ \hline
mvt                                              & \multicolumn{1}{c|}{1}                & 1.046             & \multicolumn{1}{c|}{1}              & 1.066          \\ \hline
nussinov                                         & \multicolumn{1}{c|}{2.803}            & 1.171             & \multicolumn{1}{c|}{1}              & 1.013          \\ \hline
seidel-2d                                        & \multicolumn{1}{c|}{1.0005}           & 1.0008            & \multicolumn{1}{c|}{1}              & 1.174          \\ \hline
symm                                             & \multicolumn{1}{c|}{1.0004}           & 1.06              & \multicolumn{1}{c|}{1.902}          & 1.072          \\ \hline
syr2k                                            & \multicolumn{1}{c|}{1.0009}           & 0.991             & \multicolumn{1}{c|}{1.903}          & 0.989          \\ \hline
syrk                                             & \multicolumn{1}{c|}{1.0005}           & 1.002             & \multicolumn{1}{c|}{1.903}          & 1.036          \\ \hline
trisolv                                          & \multicolumn{1}{c|}{1}                & 1.223             & \multicolumn{1}{c|}{1}              & 1.232          \\ \hline
trmm                                             & \multicolumn{1}{c|}{1.0004}           & 0.946             & \multicolumn{1}{c|}{1}              & 0.975          \\ \hline
Geomean                                          & \multicolumn{1}{c|}{\textbf{1.10193}} & \textbf{1.040725} & \multicolumn{1}{c|}{\textbf{1.383}} & \textbf{1.045} \\ \hline  
\end{tabular}
 \label{tab:res:combinedPoly}
\end{table}
 
An inspection of O3's pipeline reveals that function inlining triggers before the first instance of loop unrolling, and thus, an enhanced function inlining may benefit the subsequent passes, including loop unrolling. Note that there are a number of passes in between the two instances of our models, i.e., DCE, etc, which change the code size of an IR, and thus, the combined code size bloat does not reflect the addition of the two instances of deployments.  
The combined results are shown in Tables \ref{tab:res:combinedPoly} and \ref{tab:res:combinedCbench}. We observe a slight increase in performance benefit for both benchmarks, i.e., from 4\% to 4.5\% for Polybench and from 2\% to 2.4\% for Cbench, respectively. Note that we don't deploy ACPO-FI on Polybench as the benchmark consists of small kernels of nested loops, which might not directly benefit from an optimized inliner. However, the combined experimental result reveals that it can be done indirectly. 
LLVM's O3 pipeline has a complex behaviour when it comes to pass activation, and as mentioned earlier, several passes, including Loop Unrolling, are activated multiple times; thus, it is a hard task to quantify the benefits of each model when deployed together and the accurate measurement of partial benefits, especially when optimizations are working at different scopes, i.e., Call Graph vs Function level. 
\taco{One exceptionally good performing benchmark is \texttt{security\_sha}, which we had already optimized by 27\% (shown on Table \ref{tab:res:combinedCbench}) using ACPOFIModel, now with 50\% performance gain. There are a series of decisions made by ACPO models that we highlight an excerpt under Listing \ref{list:ACPOCombined_security_sha}. Additionally, each corresponding line at Listing \ref{list:O3_security_sha} shows the decisions made by LLVM's O3. 
Note that partial and runtime unrolling are both disabled by default at LLVM O3 and thus, \texttt{F[sha\_transform] Loop \%for.body5} is not unrolled in Listing \ref{list:O3_security_sha}. To be abundantly fair, we measured the performance of O3 when these two flags were forced enabled as well and we noticed that the decision to unroll the loop was changed to 8 as opposed to ACPOLUModel's prediction of 16 shown in Listing \ref{list:ACPOCombined_security_sha}. However, evaluations showed that still ACPO was outperforming O3 by a margin of 7\% on unrolling this particular benchmark. 
Further inspection revealed that O3's threshold mechanism built into the current version of Loop Unroll pass was limiting the assignment of higher values of unroll count even though the trip count was available at compile time. Similarly, the decision to not inline several callsites including \texttt{sha\_transform} is limited by the LLVM Inliner's threshold mechanism.} 

\begin{figure*}[!h]
\scriptsize
\noindent\begin{minipage}{.48\textwidth}
\scriptsize
\begin{lstlisting}[language=C++, label={list:ACPOCombined_security_sha}, caption=ACPO  Decisions]
//Security_sha Compilation (ACPO)
...
Loop Unroll: F[main] Loop %for.body
    Loop Size = 10 
--- "ACPOModel is activated" ---
Registering LU features: 30
Calling ML IF for inference
Final unroll type post-legality checks is: 
runtime with the unroll count of: 4
UNROLLING loop %for.body by 4!
--------------------------------------------
Loop Unroll: F[byte_reverse] Loop %for.body
 Loop Size = "15" 
--- "ACPOModel is activated" ---
Registering LU features: 30
Calling ML IF for inference
Final unroll type post-legality checks is: 
runtime with the unroll count of: 8
COMPLETELY UNROLLING loop %for.body 
with trip count 8!
...
--------------------------------------------
Loop Unroll:F[sha_transform] Loop %for.body5
  Loop Size = "20"
--- "ACPOModel is activated" ---
Registering LU features: 30
Calling ML IF for inference
    partially unrolling with count: 16
UNROLLING loop %for.body5 by 16 
with a breakout at trip 0!
--------------------------------------------
Inlining calls in: sha_update
--- "ACPOModel is activated" ---
Registering FI features: 17
Calling ML IF for inference
ACPOModel inline prediction: 1 
Updated inlining SCC: (sha_update)

--------------------------------------------
Loop Unroll: F[sha_update] Loop %while.body
  Loop Size = "457"
--- "ACPOModel is activated" ---
Registering LU features: 30
Calling ML IF for inference
runtime with the unroll count of: 16
...
--------------------------------------------
Terminating ML interface
Flushing page cache, dentries and inodes...
Done flushing.
...
Execution Time: 5.3328
\end{lstlisting}
\end{minipage}\hfill
\begin{minipage}{.48\textwidth}
\scriptsize
\begin{lstlisting}[language=C++, label={list:O3_security_sha}, caption=O3 Decisions]
//Security_sha Compilation (LLVM's O3)
...
Loop Unroll: F[main] Loop %for.body
    Loop Size = 10 
 will not try to unroll loop with runtime trip 
 count -unroll-runtime not given   
 



--------------------------------------------
Loop Unroll: F[byte_reverse] Loop %for.body
 Loop Size = "11" 
 
 
 
 
 
COMPLETELY UNROLLING loop %for.body 
with trip count 8!
...
--------------------------------------------
Loop Unroll:F[sha_transform] Loop %for.body5
  Loop Size = 15
  Trip Count = 64
  will not try to unroll partially because 
  -unroll-allow-partial not given
  "Refer to the explanations in Section 5.3"


--------------------------------------------
Inliner visiting SCC: sha_update: 2 call sites.
    Inlining (cost=285, threshold=375), Call:   
    call fastcc void @byte_reverse(i64* ...)
    NOT Inlining (cost=625, threshold=625):   
    call fastcc void @sha_transform(%struct...)
    NOT Inlining (cost=625, threshold=625):   
    call fastcc void @sha_transform(%struct...)
--------------------------------------------
Loop Unroll: F[sha_update] Loop %while.body
  Loop Size = 73
  will not try to unroll loop with runtime trip 
 count -unroll-runtime not given
}

...
--------------------------------------------

Flushing page cache, dentries and inodes...
Done flushing.
...
Execution Time: 8.012
\end{lstlisting}
\end{minipage}
\end{figure*}

\section{Discussion}
\label{sec:discussion}


ACPO is the first scalable framework for bringing ML into compilers. \taco{It provides the ability to reproduce models on different architecture very fast and the methodology lays out the steps to regenerate data, retrain a new model, and easily deploy with LLVM. Compiler autotuning and optimization space have benefited from ML for almost two decades. As we pointed out earlier, there are many recent works leveraging ML, however, the goal of ACPO is to provide a novel flexible framework for the LLVM community to leverage ML on any optimization passes of interest with minimal efforts.} Nevertheless, there are a number of challenges and shortcomings we would like to mention here.

\paragraph{Integration and dependencies} 
ACPO attempts to alleviate the challenging task of supply chain management and the use of licensed tools by separating the ML framework from the compiler. We showed both the integration of PyTorch and TensorFlow among other moving objects, i.e., Autotuner, into our framework, however, upstreaming would remain challenging and requires community efforts to shape the next generations of ML-enabled compilers. 
ACPO provides a class of features at varying LLVM scopes, therefore, new passes can utilize the already available features with minimal or no effort. It also provides the flexibility of supporting other types of features, i.e., Graph-based, embeddings, etc. Additionally, we provide the \major{Ahead-Of-Time (AOT) compilation facilities as an alternative method of inference to the developers. As mentioned in Section \ref{sec:proposed:ml_api}, ACPO uses Python named pipes to perform interprocess communication. Although this allows for greater flexibility, it adds a slight overhead to the compilation time (Shown in Table \ref{tab:compilation_time_overhead}). 
Developers can switch between these modes when building the compiler. The flexibility of choosing between the two modes allows for a fast iterative prototyping of ML models during design and training without the need to rebuild the compiler infrastructure after every change and also, an efficient compilation overhead at deployment once the ML models are optimized and ready to be shipped.}
\paragraph{Multi-objective optimization} 
ACPO optimizes the performance of a code region with intelligent unrolling decisions. However, in a number of cases, we experimentally observe that the emitted code size is also increased. This is a(n) (in)direct result of Inliner and Loop Unroll which increase the number of instructions and thus, the code size. Although this is outside the scope of the current work, an enhanced version may also benefit from taking into account both of the objectives in its exploration strategy by identifying the Pareto optimality \cite{hoste2008cole,Ashouri2013VLIW,CK-ReQuest, silva2021exploring}. Similar strategies are employed for a number of adjacent problems; i.e, Multicore Embedded Systems \cite{ascia2005dse,silvano2011multicube} and Compiler autotuning \cite{hoste2008cole,Ashouri2013VLIW,CK-ReQuest}.  


\section{Conclusion}
\label{sec:conclusion}

In this paper, we presented ACPO, a \texttt{scalable} and \texttt{modular} ML-guided compiler framework capable of optimizing performance for LLVM. We demonstrated that using the ACPO framework, we are able to enable Loop Unroll and Function Inline passes to benefit from a seamless integration of ML models, i.e., ACPOLUModel and ACPOFIModel, for which we experimentally achieved on average 4\% and \%4.5 on Polybench, and 2.1\% and 2.5\% on Cbench. ACPO provides a generalized feature collection class for which compiler developers can add to their instantiated ACPOModel class of their need and provides utility classes to use other types of features as well.


\balance
\bibliographystyle{IEEEtran}
\bibliography{acmart}

\clearpage
\onecolumn
\appendix
\renewcommand{\thesubsection}{\Alph{subsection}}
\label{sec:appendix}

\subsection{ACPO ML Interface Class}

The \texttt{ACPOMLInterface} class is virtual and meant to be implemented depending on how your particular ML framework needs are accessed. If it is a compiled library linked with your version of LLVM or a separate process (as was the case for this paper), this class abstracts the exchange of data between the compiler and the ML framework so that it is fundamentally irrelevant to how the ML framework actually works. To use this class, the \texttt{ACPOModel} invokes routines such as \texttt{loadModel}, \texttt{freeModel}, \texttt{setCustomFeatures}, and others to set up the model on the side of the ML framework. Then, when the input data is ready, \texttt{runModel} is invoked to perform inference. The result is then retrieved by the \texttt{ACPOModel} via the \texttt{getModelResult} family of methods.

\begin{figure}[h]
\scriptsize
\begin{lstlisting}[language=C++, label={list:ACPOMLInterface}, caption=ACPOMLInterface Header]
class ACPOMLInterface {
public:
  ACPOMLInterface();
  virtual ~ACPOMLInterface();
  virtual bool loadModel(std::string ModelSpecFile);
  virtual bool freeModel(std::string ModelName);
  virtual bool registerModel(std::string ModelName, int NumFeatures, int NumOutputs);
  virtual bool registerFeature(std::string ModelName, std::string FeatureName, int Index);
  virtual bool registerOutput(std::string ModelName, std::string OutputName, std::string OutputType);
  virtual bool setCustomFeature(std::string ModelName, uint64_t FeatureID, int FeatureValue);
  ...
  virtual bool setCustomFeature(std::string ModelName, std::string FeatureName, bool FeatureValue);
  virtual bool setCustomFeatures(std::string ModelName, const std::vector<std::pair<uint64_t, std::string>> &FeatureValues);
  virtual bool setCustomFeatures(std::string ModelName, const std::vector<std::pair<std::string, std::string>> &FeatureValues);
  ...
  virtual bool runModel(std::string ModelName);
  virtual std::string getOutputType(std::string ModelName, std::string OutputName);
  virtual int getModelResultI(std::string OutputName);
  ...
  virtual bool getModelResultB(std::string OutputName);
  ...
  virtual int getStatus();
  virtual bool closeMLInterface();
};
\end{lstlisting}
\end{figure}


\subsection{ACPO Loop Unrolling Model}
We start by providing the necessary arguments for the constructor and initializing the private members. Subsequently, We run a few checks to make sure the steps mentioned prior under Section \ref{sec:proposed:ml_api} are performed correctly. That leads us to the main inference method which is \texttt{getAdviceML()}. 
The texttt{getAdviceML() method first retrieves a reference to the ML interface which is used to communicate with the ML framework. It then loads a model used for inference, which causes the framework to load the appropriate model as specified in the \texttt{model.acpo} file.
If the model has already been loaded by the framework, then this command completes instantly, as loading the same model many times serves no purpose. 
The method then sets the inputs to the model and invokes the \texttt{runModel} method to begin inference.
The inference process takes place on the ML framework, and the \texttt{runModel} call is a blocking call that waits until the inference completes. When it does, a successful inference will return desired results, which can be retrieved using \texttt{getModelResultI} methods. However, if for any reason inference fails, the result of \texttt{runModel} will be false, indicating that the process failed. 
This mechanism allows the compiler to handle this case as the developer chooses. In our case, we forced the compiler to terminate with an error to ensure our ML model was the only mechanism for determining profitability. However, developers are welcome to provide fall-back mechanisms as they see fit.

\begin{figure}[h]
\footnotesize
\begin{lstlisting}[language=C++, label={list:ACPOLUModel}, caption=ACPOLUModel Instantiation]
class ACPOLUModel : public ACPOModel {
public:
  ACPOLUModel(Loop *L, OptimizationRemarkEmitter *ORE, bool UseML = true)
      : ACPOModel(ORE, UseML), CurrentLoop(L), UnrollingFactor(0), UnrollingType(0) {
        std::vector<BasicBlock *> BBVec = L->getBlocksVector();
    ...
    setMLIF(createPersistentPythonMLIF());
  }
  ...
protected:
  // Get an Advice inferred from the ML Model
  virtual std::unique_ptr<ACPOAdvice> getAdviceML() override {
    std::shared_ptr<ACPOMLInterface> MLIF = getMLIF();
    MLIF->loadModel("model-lu.acpo");
    MLIF->setCustomFeatures("LU", CustomFeatureValues);
    bool ModelRunOK = MLIF->runModel("LU");
    ...
    UnrollingType = MLIF->getModelResultI("LU-Type");
    UnrollingFactor = MLIF->getModelResultI("LU-Count");
    ...
    // Form Advice object to return to caller
    std::unique_ptr<ACPOAdvice> Advice = std::make_unique<ACPOAdvice>();
    Advice->addField("LU-Count", ConstantInt::get(Type::getInt64Ty(*(getContextPtr())),(int64_t)UnrollingFactor));
    Advice->addField("LU-Type", ConstantInt::get(Type::getInt64Ty(*(getContextPtr())),(int64_t)UnrollingType));
    return Advice;
  }
  ...
};
\end{lstlisting}
\end{figure}

\subsection{ACPO Function Inlining Model}
We instantiate an \texttt{ACPOFIModel} instance, which in itself creates a persistent ML interface instance to a Python-based ML Framework, similar to what we showed earlier with loop unroll pass. Once the \texttt{ACPOFIModel} is instantiated, we invoke setCustomFeatures() to specify features as inputs to the model. This fulfills the requirements of the \texttt{ACPOFIModel} class and ensures all inputs are provided as needed before an inference is called. The \texttt{getAdvice} method is then called, which in turn invokes \texttt{getAdviceML} which facilitates the process of inference on behalf of the compiler. The result is then decomposed into "FI-ShouldInline" field, which is then obeyed by the function inlining pass and applied as directed. Note that in this work, similar to loop unroll pass, we only replace the profitability checks with ML models.

\begin{figure}[h]
\footnotesize
\begin{lstlisting}[language=C++, label={list:ACPOFIModel}, caption=ACPOFIModel Instantiation]
class ACPOFIModel : public ACPOModel {
public:
  ACPOLFIModel(CallBase *CB, InlineAdvisor *IA, OptimizationRemarkEmitter *ORE, 
               bool OnlyMandatory, bool UseML = true)
      : ACPOModel(ORE, UseML), CurrentCB(CB), NotACPOAdvisor(IA), OnlyMandatory(OnlyMandatory) {
    Function *Caller = CB->getCaller();    ...
    setMLIF(createPersistentPythonMLIF());
  }
  ...
protected:
  // Get an Advice inferred from the ML Model
  virtual std::unique_ptr<ACPOAdvice> getAdviceML() override {
    std::shared_ptr<ACPOMLInterface> MLIF = getMLIF();
    MLIF->loadModel("model-fi.acpo");
    MLIF->setCustomFeatures("FI", CustomFeatureValues);
    bool ModelRunOK = MLIF->runModel("FI");
    ...
    ShouldInline = MLIF->getModelResultI("FI-ShouldInline");
    ...
    // Form Advice object to return to caller
    Advice->addField("FI-ShouldInline",
                     ConstantInt::get(Type::getInt64Ty(*(getContextPtr())),
                                      (int64_t)ShouldInline));
    return Advice;
  }
  ...
};
\end{lstlisting}
\end{figure}

\end{document}